\documentclass[a4paper,11pt]{article}
\usepackage{pos}

\usepackage{tikz,graphicx,booktabs,multirow,array,microtype,mathtools,float}
\graphicspath{{figures/}}
\usepackage{enumerate}  

\usepackage{anyfontsize}
\usepackage[shortlabels]{enumitem}

\usepackage{comment} 
\excludecomment{supplement}

\usepackage{amsfonts}
\AtBeginDocument{
\heavyrulewidth=.08em
\lightrulewidth=.05em
\cmidrulewidth=.03em
\belowrulesep=.65ex
\belowbottomsep=0pt
\aboverulesep=.4ex
\abovetopsep=0pt
\cmidrulesep=\doublerulesep
\cmidrulekern=.5em
\defaultaddspace=.5em
}

\usepackage{amsmath} 
\usepackage{dsfont}  
\usepackage{bm}      

\def\reals{\mathds{R}}
\def\beq{\begin{equation}}
\def\eeq{\end{equation}}
\def\beqs#1\eeqs{\beq\begin{split} #1 \end{split}\eeq}

\long\def\comment#1{}

\usepackage{mleftright,xparse}
\NewDocumentCommand\xDeclarePairedDelimiter{mmm}
{%
	\NewDocumentCommand#1{som}{%
		\IfNoValueTF{##2}
		{\IfBooleanTF{##1}{#2##3#3}{\mleft#2##3\mright#3}}
		{\mathopen{\csname##2\endcsname#2}##3\mathclose{\csname##2\endcsname#3}}%
	}%
}
\xDeclarePairedDelimiter{\av}{\langle}{\rangle}
\xDeclarePairedDelimiter{\ket}{|}{\rangle}
\xDeclarePairedDelimiter{\bra}{\langle}{|}
\NewDocumentCommand\braket{somm}{%
	\IfNoValueTF{#2}{\mleft\langle #3\,|#4\mright\rangle}{NOTIMPLEMENTED}
}
\NewDocumentCommand\opbraket{sommm}{%
	\IfNoValueTF{#2}
	{\IfBooleanTF{#1}{\langle#3|#4|#5\rangle}{\mleft\langle #3 \left| #4 \right| #5 \mright\rangle}}
	{\mathopen{\csname#2\endcsname\langle}#3\mathopen{\csname#2\endcsname|} #4 \mathclose{\csname#2\endcsname|} #5\mathclose{\csname#2\endcsname\rangle}}
}

\title{Higher order quantization conditions for two spinless particles}
\ShortTitle{Higher order quantization conditions...}

\author*[a]{Frank X. Lee}
\author[a]{Andrei Alexandru}
\author[a]{Ruair\'i Brett}

\affiliation[a]{Physics Department, The George Washington University, Washington, DC 20052, USA}

\emailAdd{fxlee@gwu.edu}
\emailAdd{aalexan@gwu.edu}
\emailAdd{rbrett@gwu.edu}

\abstract{Lattice QCD calculations of scattering phaseshifts and resonance parameters in the two-body sector are becoming precision studies. 
Early calculations employed L\"uscher's formula for extracting these quantities at lowest order.
As the calculations become more ambitious, higher-order relations are required.
In this study we derive higher-order quantization conditions and introduce a method to transparently cross-check our results. 
This is an important step given the involved derivations of these formulae.
We derive quantization conditions up to $\ell=5$ partial waves in both cubic and elongated geometries, and for states with zero and non-zero total momentum.
All 45 quantization conditions we include here (22 in cubic box, 23 in elongated box) pass our cross-check test.}

\FullConference{%
 The 38th International Symposium on Lattice Field Theory, LATTICE2021
  26th-30th July, 2021
  Zoom/Gather@Massachusetts Institute of Technology
}

\begin{document}
\maketitle

\section{Introduction}
\label{sec:QC}
The quantization condition (QC) for zero-momentum states of
two equal-mass, spinless particles in a periodic cubic box worked out by L\"uscher~\cite{Luscher:1990ux}
opened up new opportunities for studying hadron-hadron interactions. 
The L\"{u}scher method, as it is known now, is very general.  
It does not matter how the energy levels are obtained, be it in quantum mechanics, effective field theories, lattice QCD, or any other method. 
The same quantization condition applies and the results are the same up to exponentially suppressed finite-volume corrections. 
For this reason, it has become the method of choice for studying strongly-interacting systems where traditional methods 
like perturbation theory do not apply.  
In the field of nuclear and particle physics, the method has proven especially successful.
Various extensions to the method have since been made to enhance its applicability in the two-hadron sector, 
including moving frames~\cite{Rummukainen:1995vs,Doring:2012eu},
asymmetric boxes~\cite{Lee:2017igf,Pelissier:2012pi,Guo:2016zos,Guo:2018zss,Culver:2019qtx}, multiple partial waves and coupled-channel scattering~\cite{Mai:2019pqr,Brett:2018jqw,Guo:2016zos,Hansen_2012,Briceno:2014oea,Morningstar_2017}.
Significant progress towards a complete three-body scattering quantization condition has also been made in recent years, though we do not discuss it here. See Refs.~\cite{Mai:2021nul,Hansen:2019nir,Culver:2019vvu,Blanton_2020,Alexandru:2020xqf,Brett:2021wyd,Hansen:2020otl} for reviews of theoretical developments, and some first applications to three-pion and kaon scattering.

In this work we derive and validate the quantization conditions for two spinless
particles of unequal masses, rest and in moving frames, cubic and elongated geometries, and partial waves as high as $\ell=5$ (The full version is in Ref~\cite{Lee:2021kfn}).

\section{Quantization condition}
\label{sec:QC}
The quantization condition connects the infinite-volume phaseshifts with the
discrete energies of two-body states in the box~\cite{Luscher:1990ux},
\beq
\det \left [e^{2i\delta(k)} - \frac{ M(k,L) +i }{ M(k,L) -i} \right ] =0.
\label{eq:phaselat}
\eeq
The $M(k,L)$ is a hermitian matrix function of CM momentum and box size, whose explicit form is given by 
\beq
M_{l m,l' m' } = \frac{ (-1)^l }{ \eta \pi^{3/2} } \sum_{j=[l-l'|}^{l+l'} \sum_{s=-j}^{j} 
\frac{ i^j }{ q^{j+1} } Z_{js}(q^2,\eta) C_{ lm,js,l'm'},
\label{eq:mmat}
\eeq
where we have adapted it to include the $z$-elongated box geometry via $\eta$. In practical applications, the matrix is further adapted to the symmetry of box. 
The zeta function is defined by
\beq
\mathcal{Z}_{lm} (q^2,\eta) = \sum_{\widetilde{\bm n}} \frac{\mathcal{Y}_{lm}(\widetilde{\bm n})}{\widetilde{\bm n}^2-q^2},
\quad \widetilde{\bm n}= (n_x,n_y,n_z/\eta), \quad q=\frac{kL}{2\pi}.
\label{eq:zfun}
\eeq
The poles of the zeta function $\tilde{n}^2=q^2$ correspond to free-particle energies in the box. 

In group theory language, the symmetry group for states at rest is $O_h$ in cubic box,  $D_{4h}$ in $z$-elongated box.
For moving states, the symmetry is described by the so-called little groups, depending in which direction the system is moving in the fixed box. 
We will consider four distinct moving frames,  $d=(0,0,1)$,  $d=(1,1,0)$, $d=(1,1,1)$, and  $d=(0,1,2)$. 
In both cubic and $z$-elongated boxes,  $d=(0,0,1)$ has $C_{4v}$ as the little group, $d=(1,1,0)$ corresponds to $C_{2v}$, and $d=(0,1,2)$ corresponds to $C_{1v}$.
However, for $d=(1,1,1)$, the little group is $C_{3v}$ in cubic box and $C_{1v}$ in $z$-elongated box.
For moving frames, the zeta functions in Eq.\eqref{eq:zfun} need to be modified to include the boost $\bm d$,
\beq
\mathcal{Z}_{lm}(q^2,\bm d,\eta) = \sum_{ \widetilde{\bm n}\in P_{\bm d}(\eta)} \frac{\mathcal{Y}_{lm}(\widetilde{\bm n})}{\widetilde{\bm n}^2-q^2}, \;
P_{\bm d}(\eta) =\left\{\widetilde{\bm n}\in\reals^3 \mid \widetilde{\bm n}=\hat{\eta}^{-1}(\bm m-\frac{1}{2}A\, \bm d), \bm m\in \mathds{Z}^3 \right\}.
\label{eq:zfun_boost_eta}
\eeq
The factor $A$ in non-relativistic kinematics is
$
A\equiv 1+ \frac{m_2 - m_1 }{m_2 + m_1}.
$
This is to be contrasted with the relativistic version 
$A=1+ (m_1^2-m_2^2) / W^2$ where $W=\sqrt{m_1^2 + k^2} + \sqrt{m_2^2 + k^2}$ is the invariant energy of the system.
Due to lack of parity in moving frames, there is mixing between odd and even $l$ states within a given irrep. 
This means that the phaseshift formulas are generally more complicated for moving states than for the ones at rest.
One consequence is the appearance of zeta functions with odd values of $l$.

In Table~\ref{tab:all}, we give an overview of the total angular momentum content in each irrep 
(or QC), as part of a larger summary. 
It is important to realize that each QC is a single condition that couples to an infinite tower of $l$ values; 
only the lowest few are shown.
The lowest partial wave in each irrep can be computed using the energy levels in the box and if the higher partial waves can be neglected.

\section{Two-particle energies in a periodic box}
\label{sec:scheq}

To check our derivation of the quantization conditions discussed in the previous
section, we want to calculate the spectrum of the two-particle states
in a finite box with periodic boundary conditions. To make the calculation
transparent we will use a non-relativistic setup with the particles' 
interaction controlled by a rotationally invariant potential. We will
solve the problem numerically using a lattice discretization of the 
Hamiltonian and the associated Schr\"odinger equation. The results
are extrapolated to the continuum limit before comparing them to
the results of the quantization conditions.

We consider the general case $L\times L\times \eta L$ where $\eta$ is the elongation factor in the $z$-direction.
We solve the Schr\"{o}dinger equation $H\Psi=E\Psi$ in the box frame (lab frame). 
We project the problem to a new basis consisting of total momentum $\bm P$ and relative coordinates $\bm r$ in the lab frame,
\beq 
\ket{\bm P,\bm r}=\sum_{\bm m} e^{i \bm P \cdot \bm m} \ket{\bm m, \bm m+ \bm r},
\label{eq:Pr}
\eeq
where $\ket{\bm n_1, \bm n_2}$ is the ket in the position representation for two particles.

The projection leads to the reduced problem $H\psi(\bm P,\bm r)=E\psi(\bm P,\bm r)$ where the lattice Hamiltonian is given on a seven-point stencil,
\fontsize{8}{8}
\beqs
H\ket{\bm P, \bm r} &=-\frac{\hbar^2}{ 2} \sum_{\mu} \frac{-1}{ 180 a^2}  \\ 
& \Big[
-2\left( \frac{e^{3iP_\mu a} }{ m_1} + \frac{1}{ m_2} \right) \ket{\bm P,\bm r+3a\hat\mu}
+27\left( \frac{e^{2iP_\mu a} }{ m_1} + \frac{1}{ m_2} \right) \ket{\bm P, \bm r+2a\hat\mu}
-270\left( \frac{e^{iP_\mu a} }{ m_1} + \frac{1}{ m_2} \right) \ket{\bm P,\bm r+a\hat\mu}
 \\ & 
 -2\left( \frac{e^{-3iP_\mu a} }{ m_1} + \frac{1}{ m_2} \right) \ket{\bm P,\bm r-3a\hat\mu}
+27\left( \frac{e^{-2iP_mu a} }{ m_1} + \frac{1}{ m_2} \right) \ket{\bm P, \bm r-2a\hat\mu}
-270\left( \frac{e^{-iP_\mu a} }{ m_1} + \frac{1}{ m_2} \right) \ket{\bm P,\bm r-a\hat\mu}
 \\ & 
+490\left( \frac{1}{ m_1} + \frac{1}{ m_2} \right) 
\ket{\bm P,\bm r}
\Big]  + V_L(\bm r)\ket{\bm P,\bm r} + O(a^6).
\label{eq:latH7}
\eeqs
\normalsize
We need to take the continuum limit to obtain box levels from lattice levels. This is done by increasing the number of grid points and deceasing the lattice spacing simultaneously while keeping the box size fixed, 
$
\lim_{\substack{a\to 0\\ N\to \infty }} N a = L.
$

We compare this spectrum with the one derived from the quantization conditions.
Both the phaseshifts in the infinite volume and the discrete energy levels in the finite volume are independently obtained. 
They are then used in the QC to examine its efficacy. 
This check is useful because the QC is often used self-consistently to `reverse engineer'  the expected energy levels based on model parametrization of the phaseshifts, with its correctness assumed. 

\section{Infinite volume phaseshifts} 
\label{sec:phases}
\begin{figure}[!htbp]
\centering\includegraphics[scale=0.35]{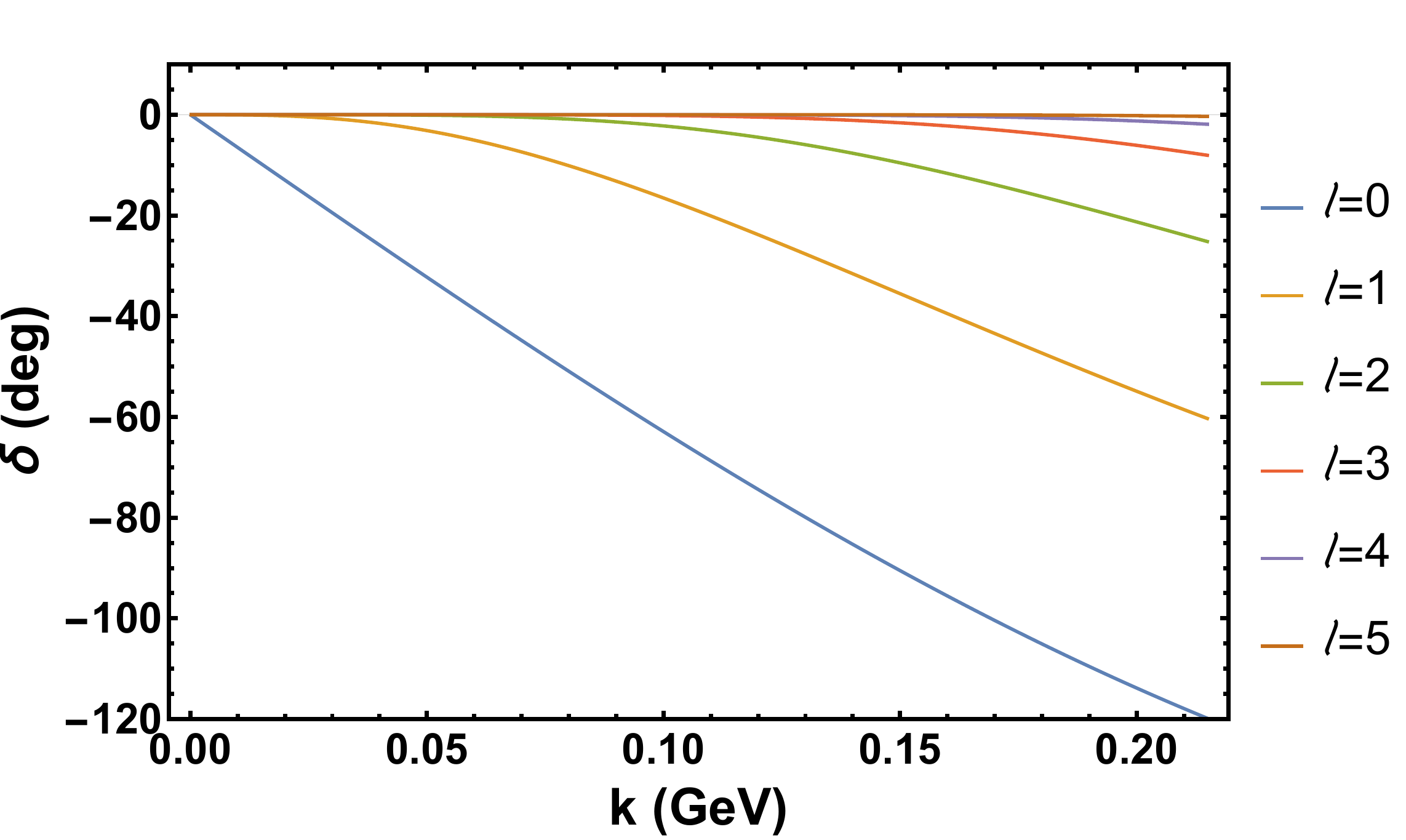}
\caption{Phaseshift of the test potential for the lowest six partial waves.}
\label{fig:phase}
\end{figure}

The first step is to compute the phase shifts for a simple potential. 
Consider two-particles $m_1=0.138$ GeV and $m_2=0.94$ GeV, interacting through a repulsive potential of Gaussian fall-off,
\beq
V(r)=C e^{-0.5(r/R_0)^2}
\label{eq:pot2}
\eeq
where  $C=1.0$ GeV and  $R_0=1.25$ fm. 
The range of the potential is about 4 fm.
The phaseshifts can be obtained readily by the variable phase method~\cite{Calogero1967}. 
For partial waves up to $l=5$ and momenta up to about 0.2 GeV, they are shown in Fig.~\ref{fig:phase}.
The phaseshifts have the expected $\delta_l(k) \sim k^{2l+1}$ asymptotic behavior. 
The potential is chosen so that in our tests partial waves up to $l=5$ 
can be checked for convergence in the $k$ range we use.
The goal is to check our derivation for the higher order QCs by comparing these energies produced by these phaseshifts with the two-particle spectrum in finite volume.
%

\section{Results and discussion} 

\begin{figure}
\centering
\includegraphics[scale=0.3]{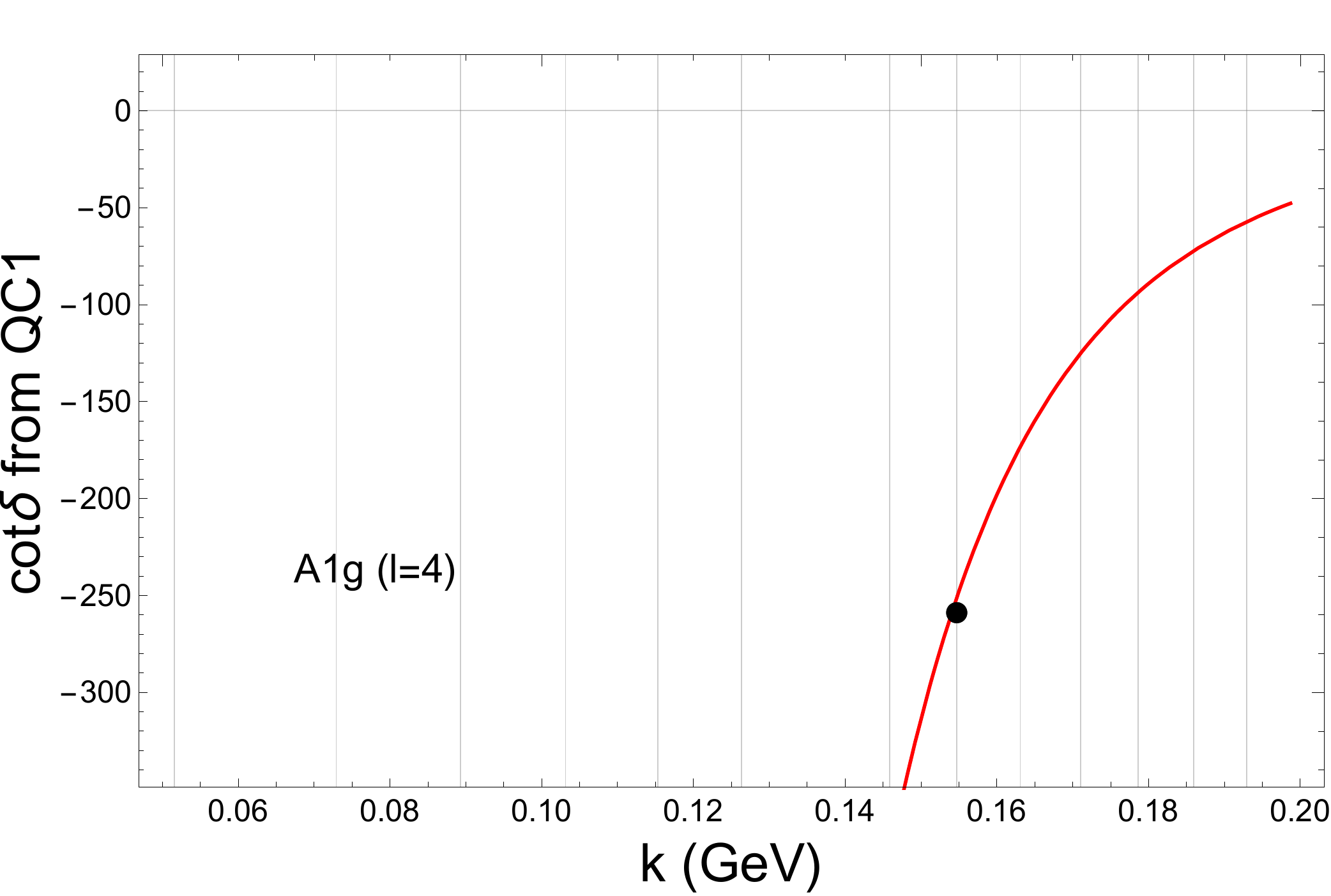}
\includegraphics[scale=0.3]{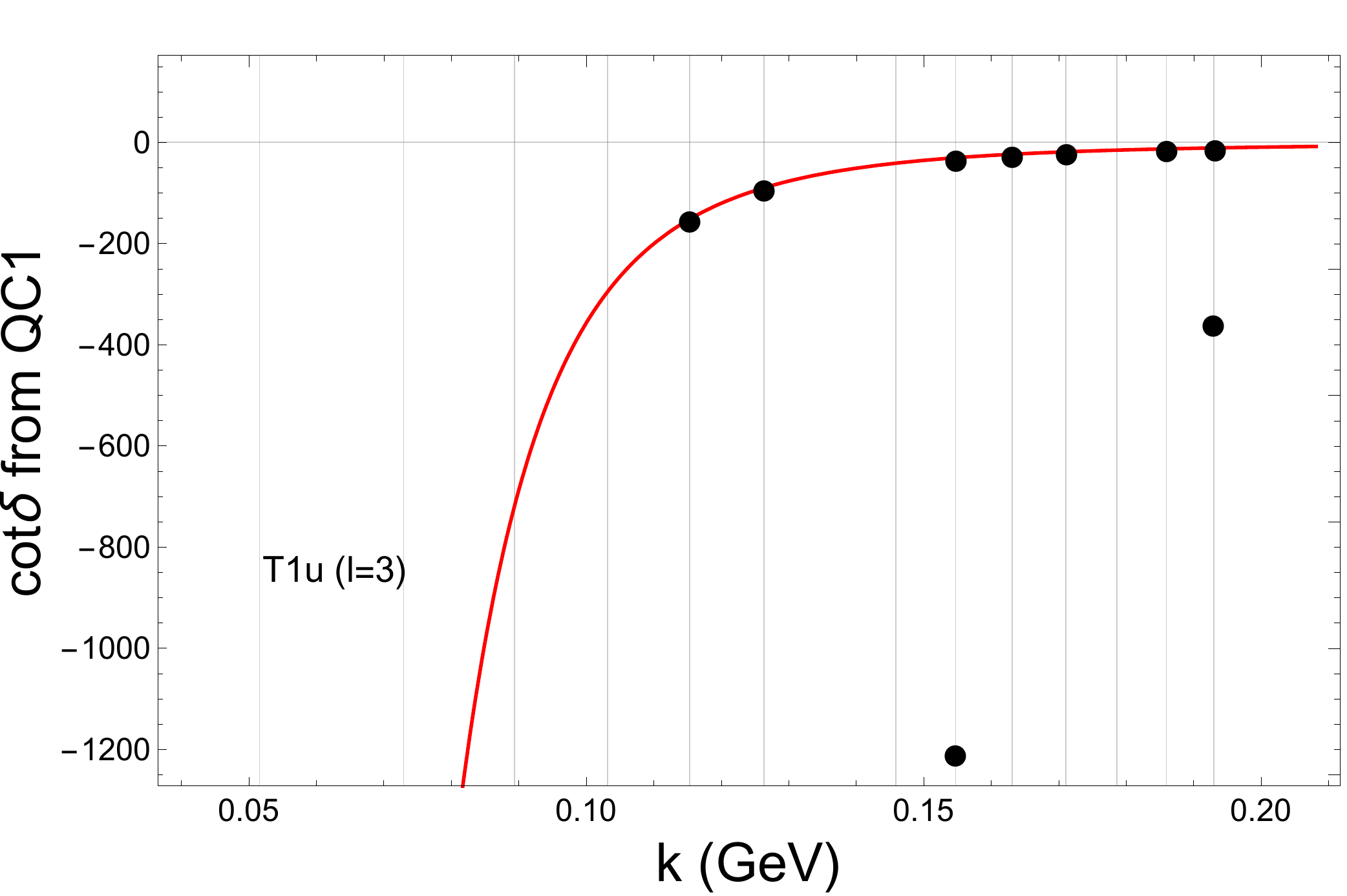}
\caption{Second partial wave in $A_{1g}$ (top) and $T_{1u}$ (bottom) of cubic box. The black points are $\cos\delta_{\rm 2nd}$ in Eq.\eqref{eq:2nd} with $\cos\delta_{\rm 1st}$ neglected, evaluated at the pinched box levels from order 1. The red curve is the infinite-volume $\cos\delta_{\rm 2nd}$. The faint vertical lines are the free-particle poles.
}
\label{fig:Oh2nd}
\end{figure}

Since the range of the potential is about 4~fm, a box size of $L=24$~fm is sufficient to make the exponential finite-volume effects negligible.
To take the continuum limit we use lattices of $20^3$ with $a=1.2$~fm, $24^3$ with $a=1$~fm, and $30^3$ with $a=0.8$~fm for cubic case. For the elongated case we use the same three lattice spacings and the same size $L=24$~fm in the $x$-, and $y$-direction but we elongate the $z$-direction by a factor of $\eta=1.5$. 
The lowest non-zero momentum in the spectrum is controlled by the box size $k_{min}=2\pi/(L\eta)$.
For the higher $k$ values the density of states gets higher. We study states with $k<0.2$ GeV. In this $k$-range only the phaseshifts for $\ell\leq 5$ are significantly different from zero (see Fig.~\ref{fig:phase}). Therefore, we expect convergence of the QCs by $l=4$ or $l=5$. 
\begin{figure}[h]
\centering
\includegraphics[width=0.48\textwidth]{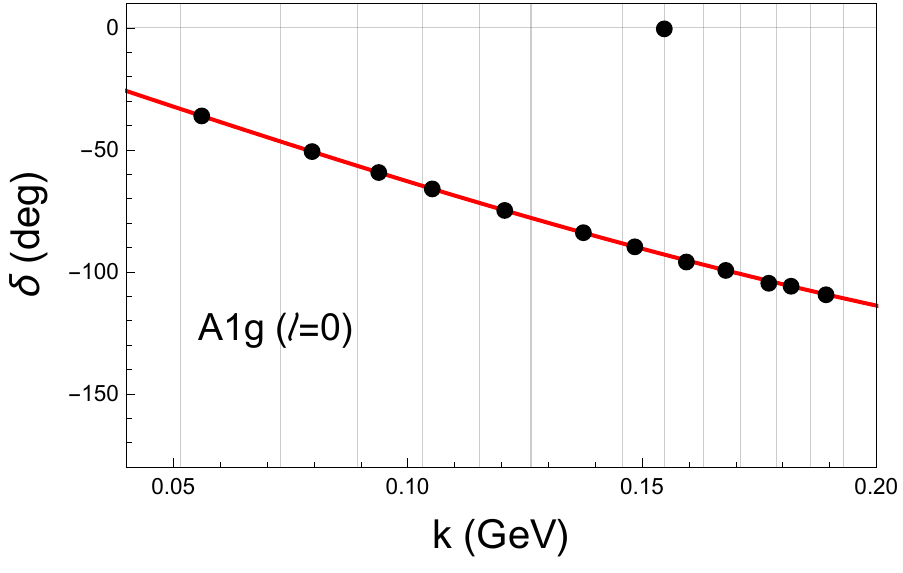}
\includegraphics[width=0.48\textwidth]{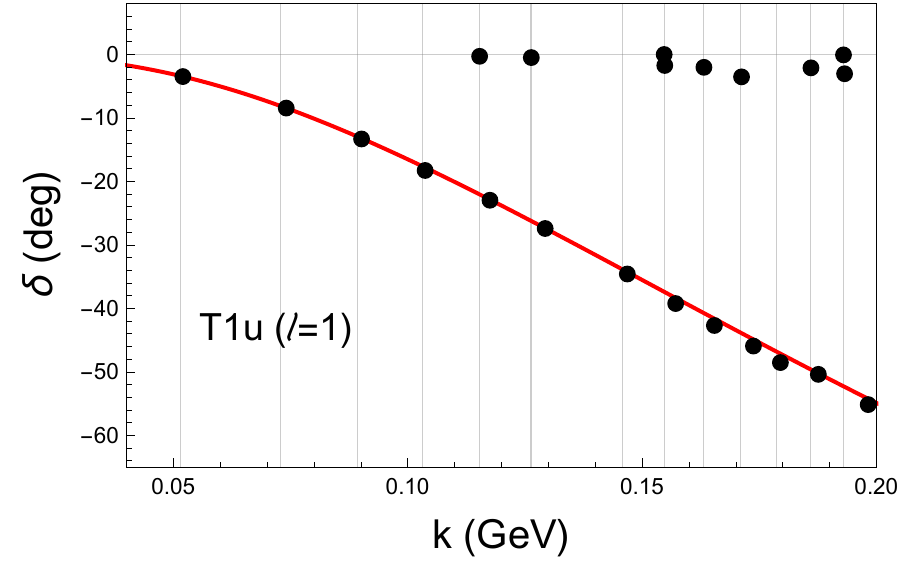}
\caption{Phaseshifts reconstruction for the lowest partial wave in the $A_{1g}$ (left) and $T_{1u}$ irreps of rest frame $d=(0,0,0)$ in cubic box. 
The black point are the predicted phaseshift via L\"uscher formula. The red curve is the infinite-volume phaseshift. 
The faint vertical lines correspond to non-interacting levels in the box. 
}
\label{fig:Oh-A1g}
\end{figure}
We show in Fig.~\ref{fig:Oh-A1g} the phaseshift prediction for the lowest $l$ in $A_{1g}$ and $T_{1u}$ by feeding the interacting energy levels into the QC. 
We see the reconstruction is excellent up to $k=0.2$~GeV, but with notable exception points.
It turns out these ``pinched'' points are sensitive to the 2nd partial waves in the QC. Solving for the second partial wave from a generic QC at order 2 with no multiplicities, we have
\beq
\cos\delta_{\rm 2nd}=M_{22}+\frac{|M_{12}|^2 }{ \cos\delta_{\rm 1st}-M_{11}}.
\label{eq:2nd}
\eeq
The ``pinched'' points occur very near the pole where $|M_{11}| \gg 1$ and
then we can approximate the equation by setting $\cos\delta_{\rm 1st}=0$
and solve for $\delta_{\rm 2nd}$.
Using above-mentioned $A_{1g}$ and $T_{1u}$ as examples, we plot in Fig.~\ref{fig:Oh2nd}  $\cos\delta_{2nd}$ extracted this way for $l=4$ and $l=3$. We see that the exception points discussed in Fig.~\ref{fig:Oh-A1g}  fall on the curve for the infinite-volume phaseshift of the 2nd partial wave.
This suggests that $\delta_{1st}(k)$ and $\delta_{2nd}(k)$ can be separately isolated by considering the QC at order 1 and order 2 respectively. 

As a general method to assess the effects of higher partial waves, we investigate the convergence of the QC by feeding it the infinite-volume phaseshifts and comparing the resulting levels with the box levels. 
We check the convergence order by order: `order 1' has only the lowest partial wave, `order 2' with the next partial wave added, and so on. In the limit that all the partial waves are included, perfect agreement is expected. 
The comparison involves very small differences that are not easily discernible visually. 
To better gauge the quality of the convergence, we introduce a numerical measure 
\beq
\chi^2=\frac{(k_{\rm box} - k_{\rm QC})^2 }{ (k_{\rm box} - k_{\rm lat})^2},
\label{eq:chi2}
\eeq
where $k_{\rm box}$ is the continuum box level extrapolated from the three lattice spacings, $k_{\rm lat}$ the level on the lattice with the finest lattice spacing, 
and $k_{\rm QC}$ the solution from the QC at each order.  The extrapolation is a linear function of $a^6$, the error present 
in the Hamiltonian from the 7-stencil approximation in Eq.\eqref{eq:latH7}. 
Basically, the convergence is measured against the tiny difference 
between the continuum box levels and those from the largest lattice used in the extrapolation (about 6 decimal places, or 1~eV out of 1~MeV). Note that the $\chi^2$ introduced is not in the standard sense of curve-fitting where the best value is around 1. Here the smaller its value, the better the convergence. 
This is a highly sensitive measure: non-convergence of a single level will have a large contribution to the total $\chi^2$.

We have confirmed the convergence of all 45 cases in the same manner, as summarized in Table~\ref{tab:all}.
All 45 QCs can be found in Ref~\cite{Lee:2021kfn}.
One point to emphasize is that to get agreement for certain cases we needed to consider QC all the way to order 5. The irreps in question (No. 10, 14, 18, 21, 33, 38, 42, 44) are, as expected, the ones that allow the most mixing between partial waves.

\begin{table*}
\caption{Summary of the total $\chi^2$-measure showing convergence for all QCs discussed in this work. 
Here $l(n)$ indicates the lowest few partial waves (and multiplicities) that couple to the QC; $N$ is the number of levels under the cutoff of $k=0.2$ GeV or 40 levels. 
Detailed convergence data for every individual energy level in Ref~\cite{Lee:2021kfn}.
}
\label{tab:all}
\centering
\begin{tabular}{c}
\renewcommand{\arraystretch}{0.8}
\fontsize{7}{7}
$
\begin{array}{cccclllllll}\toprule     
\multicolumn{11}{c}{\text{\bf Cubic box}}  \\

\text{No.}  & d & \text{Group}    & \text{QC} & l(n)  & \text{N}  & \text{Order 1} & \text{Order 2}  &\text{Order 3}  & \text{Order 4}  & \text{Order 5}   \\
\hline
1  & (0,0,0) & O_h  & A_{1g} & 0, 4, 6, \cdots & 14 & 277.206 & 3.7296 & & &\\
    &             &         & A_{1u} & 9, 13, 15 , \cdots & & &&& &\\
    &             &         & A_{2g} & 6, 10, 12, \cdots  & & &&& &\\
2  &             &         & A_{2u} & 3, 7, 9, \cdots & 6 & 0.0458813  && & &\\
3  &             &         & E_{g} &  2, 4, 6, \cdots & 16 & 2123.07 & 5.44639 & & &\\
4  &             &         & E_{u} & 5, 7, 9, \cdots & 5 & 0.0546959 & & &\\
5  &             &         & T_{1g} &4, 6, 8(2), \cdots &9 &  0.243897 && & &\\
6  &             &         & T_{1u} & 1, 3, 5(2), \cdots & 22 & 67276.5 & 36.5936 & 3.35684 & &\\
7  &             &         & T_{2g} & 2, 4, 6(2), \cdots & 16 &  2485.36 & 2.70125 & & &\\
8  &             &         & T_{2u} & 3, 5, 7(2), \cdots & 14 &  4.55575 & 0.267262 & & &\\

10  & (0,0,1) & C_{4v}  & A_1 & 0, 1, 2, 3, 4(2), \cdots & 40  &  4.08008\times 10^8 & 1.06106\times 10^{10} & 10175.7 & 284.022 & \
6.92792 \\
11  &             &             & A_2 & 4, 5, 6, \cdots & 15 &  7.46202 & 0.145375 & &\\
12  &             &             & B_1 & 2, 3, 4, 5, \cdots & 34 & 276889. & 402.025 & 5.9668 & &\\
13 &             &             & B_2 &  2, 3, 4, 5, \cdots & 27 &  350956. & 779.183 & 3.84012 & &\\
14 &             &             & E &  1, 2, 3(2), 4(2), \cdots & 40 & 1.32751\times 10^7 & 72379.5 & 514.979 & 8.21687  &\\

14  & (1,1,0) & C_{2v}  & A_1 & 0, 1, 2(2), 3(2), 4(3), \cdots &  40 & 1.96654\times 10^8 & 4.77126\times 10^6 & 17298.1 & 190.94 & 5.34113 \\
15  &             &             & A_2 & 2, 3, 4(2), 5(2), \cdots & 40 & 54211.3 & 2998.61 & 3.01327& &\\
16  &             &             & B_1 & 1, 2, 3(2), 4(2), \cdots & 40 &  2.84806\times 10^6 & 47754.8 & 241.517 & 4.62961 &\\
17 &             &             & B_2 &  1, 2, 3(2), 4(2), \cdots & 40 & 9.36978\times 10^6 & 88719.7 & 308.013 & 6.28652  &\\

18  & (1,1,1) & C_{3v}  & A_1 &  0, 1, 2, 3(2), 4(2), \cdots & 40 & 9.16556\times 10^7 & 1.09895\times 10^6 & 22260. & 115.412 & 3.53446 \\
19  &             &             & A_2 & 3, 4, 5, \cdots & 26 & 111.322 & 0.44229 & & &\\
20  &             &             & E & 1, 2(2), 3(2), 4(3), \cdots & 40 & 8.06257\times 10^6 & 23959. & 271.289 & 4.13508&\\

21  & (0,1,2) & C_{1v}  & A_1 &  0, 1(2), 2(3), 3(4), 4(5), \cdots & 40 &  1.01852\times 10^8 & 993629. & 6858.3 & 17.3429 & 2.01417 \\
22  &            &              & A_2 &  1, 2(2), 3(3), 4(4), \cdots & 40 & 8.98645\times 10^6 & 37314.8 & 293.484 & 3.18387  &\\

\hline   
\multicolumn{11}{c}{\text{\bf Elongated box}}   \\
  
\text{No.}  & d & \text{Group}    & \text{QC} & l(n)  & \text{N}  & \text{Order 1} & \text{Order 2}  &\text{Order 3}  & \text{Order 4}  & \text{Order 5}   \\
\hline
23  & (0,0,0) & D_{4h}  & A_{1g} &0, 2, 4(2), 5, \cdots & 40  & 4.06688\times 10^6 & 573.449 & 8.64225  & &\\
24  &             &         & A_{1u} & 5, 7, 9(2) , \cdots & 9  & 0.134224  & &  & &\\
25    &             &         & A_{2g} & 4, 6, 8(2) , \cdots & 3  &  0.00828583 & &  & &\\
26    &             &         & A_{2u} & 1, 3, 5(2), 7(2), \cdots & 34  & 131946. & 16.2795 & 10.7998 & &\\
27  &             &         & E_{g} & 2, 4(2), 6(3), \cdots & 38  &  2009.93 & 0.319341 & & &\\
28  &             &         & E_{u} & 1,3(2),5(3), \cdots & 40  &   221693. & 41.2595 & 8.52709  & &\\
29  &             &         & B_{1g} & 2, 4, 6(2), 8(2),\cdots & 27  & 703.291 & 11.0395 & & &\\
30  &             &         & B_{1u} & 3,5,7(2), 9(2), \cdots & 17  &  7.07159 & 0.168506 & & &\\
31  &             &         & B_{2g} & 2, 4, 6(2), 8(2), \cdots & 22  &  2093.46 & 0.192 & & &\\
32  &             &         & B_{2u} & 3, 5, 7(2), 9(2), \cdots & 21  & 2.06599 & 0.488025 & & &\\

33  & (0,0,1) & C_{4v}  & A_1 & 0, 1, 2, 3, 4(2), \cdots & 40  &  5.70458\times 10^8 & 9.64404\times 10^6 & 14025.4 & 166.777 & \
11.0137 \\
34  &             &             & A_2 & 4, 5, 6, \cdots & 21 & 40.092 & 0.21509 & &\\
35  &             &             & B_1 & 2, 3, 4, 5, \cdots & 40 & 161200. & 274.945 & 5.1077 & &\\
36 &             &             & B_2 &  2, 3, 4, 5, \cdots & 39 & 318459. & 845.868 & 2.82798 & &\\
37 &             &             & E &  1, 2, 3(2), 4(2), \cdots & 40 & 1.66366\times 10^7 & 79648.3 & 353.966 & 7.28402 &\\

38  & (1,1,0) & C_{2v}  & A_1 & 0, 1, 2(2), 3(2), 4(3), \cdots & 40 & 2.19334\times 10^8 & 3.60527\times 10^6 & 9669.56 & 99.3024 & 27.4093\\
39  &             &             & A_2 & 2, 3, 4(2), 5(2), \cdots & 40 & 34211.4 & 295.311 & 0.804797& &\\
40  &             &             & B_1 & 1, 2, 3(2), 4(2), \cdots & 40 & 2.97904\times 10^6 & 28179. & 115.601 & 6.80754  &\\
41 &             &             & B_2 &  1, 2, 3(2), 4(2),  \cdots & 40 & 9.05592\times 10^6 & 71558.8 & 146.885 & 18.7042  &\\

42  & (1,1,1) & C_{1v}  & A_1 &  0, 1(2), 2(3), 3(4), 4(5), \cdots & 40  & 5.45564\times 10^8 & 2.89417\times 10^6 & 15983.3 & 39.1208 & 8.00598 \\
43  &             &             & A_2 & 1, 2(2), 3(3), 4(4), \cdots & 40  &  6.85779\times 10^6 & 23813.9 & 69.434 & 3.24328 &\\

44 & (0,1,2) & C_{1v}  & A_1 &  0, 1(2), 2(3), 3(4), 4(5), \cdots & 40  & 3.79138\times 10^8 & 3.16621\times 10^6 & 6213.69 & 17.0296 & 5.34524\\
45  &            &              & A_2 &  1, 2(2), 3(3), 4(4), \cdots & 40  &  1.27375\times 10^7 & 45256.8 & 146.374 & 2.67257 &\\

\bottomrule
\end{array}
$     
\end{tabular}
\end{table*}
\normalsize

\section{Conclusion and outlook}
\label{sec:con}

We derived higher-order L\"uscher quantization conditions (QC) for scattering of two spinless particles of unequal masses.
Our results were checked numerically by comparing the QC predictions with the spectrum of two-particle states in a box computed
by solving the Schr\"odinger equation.
This is done using a simple potential model in non-relativistic quantum mechanics. Both the phaseshifts in infinite volume and energy levels in finite volume are independently generated in a well-controlled fashion. 
Here is a summary of our findings.

\begin{enumerate}[1)]
\item
We considered a variety of scenarios: rest frame and four moving frames, cubic and elongated geometries. In total, we examined 22 QCs in the cubic box and 23 QCs in the elongated box. 
The five lowest partial waves in each QC are examined. In some cases, up to $l=5$. 
 Some of the QCs are re-derived to include higher partial waves, others are new.
Generically, we expect the QCs to be valid up to terms which vanish exponentially with the box size.

\item
We choose the potential and the box-size so that the systematics associated with finite-volume are negligible, on one hand,
and on the other the results are sensitive to partial-waves as high as $\ell=5$. This allows us to
provide very stringent tests for our results.
The numerical checks are done at high precision (to six decimal digits, or differences of 1 eV resolved out of 1 MeV). 
Up to CM momentum $k=0.2$ GeV and up to 40 levels are examined for each of the QCs.

\item
We found sensitivity to the second lowest partial wave in selected QCs through `pinched' levels which coincide with free-particle poles. The sensitivity can be used to provide an approximate phaseshift for the second lowest partial wave despite the presence of the lowest one in a particular channel, but this must be determined  on a case by case basis.
If such levels are encountered in lattice QCD simulations, they can be either ignored or used to estimate the second partial wave.

\item
For the most part, we find elongated boxes work just as well as cubic ones.  This bodes well for using elongated boxes 
as a cost-effective way of varying the kinematic range with a modest increase in the lattice volume. 

\item
Boosting of the two-particle system in both the cubic and elongated boxes allows lower energies to be accessed, thus a wider coverage. 
The trade-off is the loss of parity which means more mixing of partial waves. 

\item
The effort is already paying dividends. For example, we checked the integer-$J$ QCs in Ref.~\cite{Gockeler:2012yj} for $d=(1,1,0)$ and $d=(1,1,1)$ and found agreement with ours, despite having different forms due to different basis vectors. Those QCs are only given for up to $l=2$. Here we extend up to $l=4$. We also checked against $C_{3v}$ up to $l=4$ from an independent source~\cite{Colin2021} and found agreement.
We also found a few typos in the QCs included in Ref.~\cite{Lee:2017igf}. 
We also checked against all the expressions up to $l=4$ for spinless particles of equal mass at total zero momentum in non-elongated boxes given in Ref.~\cite{Luscher:1990ux} by setting $m_1=m_2$ in our expressions and found agreement.

\end{enumerate}

For outlook, we envision the following possibilities.
\begin{enumerate}[1)]
\item
The QCs can only be used to extract phaseshifts from energy only for the lowest partial waves in each irrep. The predictions are affected by 
cutting off all the higherr partial waves. The severity is not known {\it a priori} and it depends on the box geometry and the total momentum of the state. The problem can be turned on its head: can we extract the higher partial waves by 
considering multiple QCs simultaneously?  We have seen in limited cases that higher partial waves can be isolated in a single QC despite the presence of a lower one. Is there a systematic approach, taking advantage of multiple irreps, moving frames, and box size? 

\item
 We note that the same methodology could be easily applied for other potentials, if there is a physical problem that requires calculation of the two-particle spectrum in a finite box.
Any interaction potential can be used in this approach, including potentials given in numerical form or nonlocal potentials $V(r, \bm p)$ where  $\bm p$ can be treated as finite differences on the lattice.

\item
The formalism can be used to study the finite-volume effects in lattice QCD simulation of physical systems, 
such as the the magnitude of exponential
finite-volume effects ignored by the QC's by considering a smaller box (3.5 to 6 fm); the effect of the range of the model potential; and/or the finite-volume spectrum in the presence
of shallow bound states. Even using the naive $O(a^2)$ discretization to study the influence of cutoff effects on the extracted finite-volume energies could be interesting.

\item
The formalism can be applied with minimal modification to systems with two integer spins, such as $\pi\rho$ or  $\rho\rho$ scattering.  The same is true for two spin-1/2 particles, such as nucleon-nucleon scattering. The spins and orbital angular momentum couple to an integer total angular momentum $J$, making such systems essentially `meson-like'. There is a plethora of NN interaction potentials to work with. It would be interesting to investigate coupled channels in such systems.

\item
Another direction is the extension to systems with total half-integer $J$, such as a spin-0 particle and a spin-1/2  particle (a classic example being the delta resonance in pion-nucleon scattering). Group theory for double-cover groups are involved for half-integer total angular momentum. 
The QCs for half-integer $J$ should also be checked since they are even more involved than the ones for integer spin. 
For this case the Hamiltonian must be modified to include spin-orbit coupling, 
such as the Fourier basis approach~\cite{Lee:2020fbo}. 
\end{enumerate}

We thank Colin Morningstar for helpful communications.
This work is supported in part by the U.S. Department of Energy grant DE-FG02-95ER40907. 

\bibliographystyle{JHEP}
\bibliography{xvalidation}

\providecommand{\href}[2]{#2}\begingroup\raggedright\begin{thebibliography}{10}

\bibitem{Luscher:1990ux}
M.~L{\"u}scher, \emph{{Two particle states on a torus and their relation to the
  scattering matrix}},
  \href{https://doi.org/10.1016/0550-3213(91)90366-6}{\emph{Nucl.Phys.}
  {\bfseries B354} (1991) 531}.

\bibitem{Rummukainen:1995vs}
K.~Rummukainen and S.A.~Gottlieb, \emph{{Resonance scattering phase shifts on a
  nonrest frame lattice}},
  \href{https://doi.org/10.1016/0550-3213(95)00313-H}{\emph{Nucl. Phys.}
  {\bfseries B450} (1995) 397}
  [\href{https://arxiv.org/abs/hep-lat/9503028}{{\ttfamily hep-lat/9503028}}].

\bibitem{Doring:2012eu}
M.~Doring, U.G.~Meissner, E.~Oset and A.~Rusetsky, \emph{{Scalar mesons moving
  in a finite volume and the role of partial wave mixing}},
  \href{https://doi.org/10.1140/epja/i2012-12114-6}{\emph{Eur. Phys. J. A}
  {\bfseries 48} (2012) 114} [\href{https://arxiv.org/abs/1205.4838}{{\ttfamily
  1205.4838}}].

\bibitem{Lee:2017igf}
F.X.~Lee and A.~Alexandru, \emph{{Scattering phase-shift formulas for mesons
  and baryons in elongated boxes}},
  \href{https://doi.org/10.1103/PhysRevD.96.054508}{\emph{Phys. Rev. D}
  {\bfseries 96} (2017) 054508}
  [\href{https://arxiv.org/abs/1706.00262}{{\ttfamily 1706.00262}}].

\bibitem{Pelissier:2012pi}
C.~Pelissier and A.~Alexandru, \emph{{Resonance parameters of the rho-meson
  from asymmetrical lattices}},
  \href{https://doi.org/10.1103/PhysRevD.87.014503}{\emph{Phys.Rev.} {\bfseries
  D87} (2013) 014503} [\href{https://arxiv.org/abs/1211.0092}{{\ttfamily
  1211.0092}}].

\bibitem{Guo:2016zos}
D.~Guo, A.~Alexandru, R.~Molina and M.~D{\"o}ring, \emph{{Rho resonance
  parameters from lattice QCD}},
  \href{https://doi.org/10.1103/PhysRevD.94.034501}{\emph{Phys. Rev.}
  {\bfseries D94} (2016) 034501}
  [\href{https://arxiv.org/abs/1605.03993}{{\ttfamily 1605.03993}}].

\bibitem{Guo:2018zss}
D.~Guo, A.~Alexandru, R.~Molina, M.~Mai and M.~D{\"o}ring, \emph{{Extraction of
  isoscalar $\pi\pi$ phase-shifts from lattice QCD}},
  \href{https://doi.org/10.1103/PhysRevD.98.014507}{\emph{Phys. Rev.}
  {\bfseries D98} (2018) 014507}
  [\href{https://arxiv.org/abs/1803.02897}{{\ttfamily 1803.02897}}].

\bibitem{Culver:2019qtx}
C.~Culver, M.~Mai, A.~Alexandru, M.~D\"oring and F.X.~Lee, \emph{{Pion
  scattering in the isospin $I=2$ channel from elongated lattices}},
  \href{https://doi.org/10.1103/PhysRevD.100.034509}{\emph{Phys. Rev. D}
  {\bfseries 100} (2019) 034509}
  [\href{https://arxiv.org/abs/1905.10202}{{\ttfamily 1905.10202}}].

\bibitem{Mai:2019pqr}
M.~Mai, C.~Culver, A.~Alexandru, M.~D\"oring and F.X.~Lee, \emph{{Cross-channel
  study of pion scattering from lattice QCD}},
  \href{https://doi.org/10.1103/PhysRevD.100.114514}{\emph{Phys. Rev. D}
  {\bfseries 100} (2019) 114514}
  [\href{https://arxiv.org/abs/1908.01847}{{\ttfamily 1908.01847}}].

\bibitem{Brett:2018jqw}
R.~Brett, J.~Bulava, J.~Fallica, A.~Hanlon, B.~H{\"o}rz and C.~Morningstar,
  \emph{{Determination of $s$- and $p$-wave $I=1/2$ $K\pi$ scattering
  amplitudes in $N_{\mathrm{f}}=2+1$ lattice QCD}},
  \href{https://doi.org/10.1016/j.nuclphysb.2018.05.008}{\emph{Nucl. Phys.}
  {\bfseries B932} (2018) 29}
  [\href{https://arxiv.org/abs/1802.03100}{{\ttfamily 1802.03100}}].

\bibitem{Hansen_2012}
M.T.~Hansen and S.R.~Sharpe, \emph{Multiple-channel generalization of
  lellouch-l{\"u}scher formula},
  \href{https://doi.org/10.1103/physrevd.86.016007}{\emph{Physical Review D}
  {\bfseries 86} (2012) }.

\bibitem{Briceno:2014oea}
R.A.~Briceno, \emph{{Two-particle multichannel systems in a finite volume with
  arbitrary spin}},
  \href{https://doi.org/10.1103/PhysRevD.89.074507}{\emph{Phys. Rev.}
  {\bfseries D89} (2014) 074507}
  [\href{https://arxiv.org/abs/1401.3312}{{\ttfamily 1401.3312}}].

\bibitem{Morningstar_2017}
C.~Morningstar, J.~Bulava, B.~Singha, R.~Brett, J.~Fallica, A.~Hanlon et~al.,
  \emph{{Estimating the two-particle $K$-matrix for multiple partial waves and
  decay channels from finite-volume energies}},
  \href{https://doi.org/10.1016/j.nuclphysb.2017.09.014}{\emph{Nucl. Phys. B}
  {\bfseries 924} (2017) 477}
  [\href{https://arxiv.org/abs/1707.05817}{{\ttfamily 1707.05817}}].

\bibitem{Mai:2021nul}
M.~Mai, A.~Alexandru, R.~Brett, C.~Culver, M.~D\"oring, F.X.~Lee et~al.,
  \emph{{Three-body dynamics of the $a_1(1260)$ resonance from lattice QCD}},
  \href{https://arxiv.org/abs/2107.03973}{{\ttfamily 2107.03973}}.

\bibitem{Hansen:2019nir}
M.T.~Hansen and S.R.~Sharpe, \emph{{Lattice QCD and Three-particle Decays of
  Resonances}},
  \href{https://doi.org/10.1146/annurev-nucl-101918-023723}{\emph{Ann. Rev.
  Nucl. Part. Sci.} {\bfseries 69} (2019) 65}
  [\href{https://arxiv.org/abs/1901.00483}{{\ttfamily 1901.00483}}].

\bibitem{Culver:2019vvu}
C.~Culver, M.~Mai, R.~Brett, A.~Alexandru and M.~D\"oring, \emph{{Three pion
  spectrum in the $I=3$ channel from lattice QCD}},
  \href{https://doi.org/10.1103/PhysRevD.101.114507}{\emph{Phys. Rev. D}
  {\bfseries 101} (2020) 114507}
  [\href{https://arxiv.org/abs/1911.09047}{{\ttfamily 1911.09047}}].

\bibitem{Blanton_2020}
T.D.~Blanton, F.~Romero-L{\'o}pez and S.R.~Sharpe, \emph{I=3 three-pion
  scattering amplitude from lattice qcd},
  \href{https://doi.org/10.1103/physrevlett.124.032001}{\emph{Physical Review
  Letters} {\bfseries 124} (2020) }.

\bibitem{Alexandru:2020xqf}
A.~Alexandru, R.~Brett, C.~Culver, M.~D\"oring, D.~Guo, F.X.~Lee et~al.,
  \emph{{Finite-volume energy spectrum of the $K^-K^-K^-$ system}},
  \href{https://doi.org/10.1103/PhysRevD.102.114523}{\emph{Phys. Rev. D}
  {\bfseries 102} (2020) 114523}
  [\href{https://arxiv.org/abs/2009.12358}{{\ttfamily 2009.12358}}].

\bibitem{Brett:2021wyd}
R.~Brett, C.~Culver, M.~Mai, A.~Alexandru, M.~D\"oring and F.X.~Lee,
  \emph{{Three-body interactions from the finite-volume QCD spectrum}},
  \href{https://doi.org/10.1103/PhysRevD.104.014501}{\emph{Phys. Rev. D}
  {\bfseries 104} (2021) 014501}
  [\href{https://arxiv.org/abs/2101.06144}{{\ttfamily 2101.06144}}].

\bibitem{Hansen:2020otl}
{\scshape Hadron Spectrum} collaboration, \emph{{Energy-Dependent $\pi^+ \pi^+
  \pi^+$ Scattering Amplitude from QCD}},
  \href{https://doi.org/10.1103/PhysRevLett.126.012001}{\emph{Phys. Rev. Lett.}
  {\bfseries 126} (2021) 012001}
  [\href{https://arxiv.org/abs/2009.04931}{{\ttfamily 2009.04931}}].

\bibitem{Lee:2021kfn}
F.X.~Lee, A.~Alexandru and R.~Brett, \emph{{Validation of the finite-volume
  quantization condition for two spinless particles}},
  \href{https://arxiv.org/abs/2107.04430}{{\ttfamily 2107.04430}}.

\bibitem{Calogero1967}
F.~Calogero, \emph{Variable phase approach to potential scattering}, Academic
  Press, New York (1967).

\bibitem{Gockeler:2012yj}
M.~Gockeler, R.~Horsley, M.~Lage, U.G.~Meissner, P.E.L.~Rakow, A.~Rusetsky
  et~al., \emph{{Scattering phases for meson and baryon resonances on general
  moving-frame lattices}},
  \href{https://doi.org/10.1103/PhysRevD.86.094513}{\emph{Phys. Rev.}
  {\bfseries D86} (2012) 094513}
  [\href{https://arxiv.org/abs/1206.4141}{{\ttfamily 1206.4141}}].

\bibitem{Colin2021}
C.~Morningstar, ``{Box matrix elements for $C_{3v}$}.'' private communication,
  2021.

\bibitem{Lee:2020fbo}
F.X.~Lee, C.~Morningstar and A.~Alexandru, \emph{{Energy spectrum of
  two-particle scattering in a periodic box}},
  \href{https://doi.org/10.1142/S0129183120501314}{\emph{Int. J. Mod. Phys. C}
  {\bfseries 31} (2020) 2050131}.

\end{thebibliography}\endgroup
\end{document}